\documentclass{article}

\usepackage{arxiv}

\usepackage[utf8]{inputenc} 
\usepackage[T1]{fontenc}    
\usepackage{hyperref}       
\usepackage{url}            
\usepackage{booktabs}       
\usepackage{amsfonts}       
\usepackage{nicefrac}       
\usepackage{microtype}      
\usepackage{lipsum}
\usepackage{subfiles}
\usepackage{graphicx}
\usepackage{xargs}          
\graphicspath{ {./} }

\usepackage{soul}
\usepackage{color, colortbl}

\usepackage{amsmath}
\numberwithin{equation}{section}

\usepackage{multirow}
\usepackage{pdflscape}
\usepackage{float}

\usepackage[colorinlistoftodos,prependcaption,textsize=tiny]{todonotes}
\newcommandx{\unsure}[2][1=]{\todo[linecolor=red,backgroundcolor=red!25,bordercolor=red,#1]{#2}}
\newcommandx{\change}[2][1=]{\todo[linecolor=blue,backgroundcolor=blue!25,bordercolor=blue,#1]{#2}}
\newcommandx{\info}[2][1=]{\todo[linecolor=olive,backgroundcolor=olive!25,bordercolor=olive,#1]{#2}}
\newcommandx{\improvement}[2][1=]{\todo[linecolor=plum,backgroundcolor=plum!25,bordercolor=plum,#1]{#2}}

\usepackage{cleveref}  

\usepackage{caption}
\usepackage{subcaption}

\definecolor{highlightTable}{gray}{0.9}
\definecolor{plum}{rgb}{0.56, 0.27, 0.52}

\newcommand{\makeAuthor}[4]{#1\\#2\\#3\\\texttt{#4}\\}
\newcommand{\makeKeywords}{
Virtual reality \and
Intuitive physics
}

\title{Human causal perception in a cube-stacking task}

\author{
 \makeAuthor{Nikolai Bahr}{Cognitive Neuroinformatics, University of Bremen}{Bremen, Germany}{nibahr@uni-bremen.de}
 \And
 \makeAuthor{Christoph Zetzsche}{Cognitive Neuroinformatics, University of Bremen}{Bremen, Germany}{zetzsche@informatik.uni-bremen.de}
}

\begin{document}
\maketitle

\begin{abstract}

In intuitive physics the process of stacking cubes has become a paradigmatic, canonical task. Even though it gets employed in various shades and complexities, the very fundamental setting with two cubes has not been thoroughly investigated. Furthermore, the majority of settings feature only a reduced, one dimensional (1D) decision space. In this paper an experiment is conducted in which participants judge the stability of two cubes stacked on top of each other. It is performed in the full 3D setting which features a 2D decision surface. The analysis yield a shape of a rotated square for the perceived stability area instead of the commonly reported safety margin in 1D. This implies a more complex decision behavior in human than previously assumed.

\end{abstract}


\section{Introduction}

In everyday life it is crucial to have some understanding of physical properties and processes for even seemingly simple tasks like cutting food, setting the table or even placing a cup onto a saucer. For this somewhat intuitive physics it might be beneficial to look at how humans behave.
Additionally, when human and robot work together, it is crucial for the robot to adopt these intuitive human heuristics. Otherwise, humans may become uncomfortable and feel the need to correct the robot's actions, hindering the development of a productive co-working environment. 

Human intuitive physics is an active area of research. In this field, the paradigmatic (\cite{conwell2021sharedParadigmatic}) task of stacking cubes - or assessing their stability - has been utilized in various forms and levels of complexity. This includes virtual reality (VR), towers with a lot of cubes or even different shapes. In spite of this, the most basic setting of this canonical (\cite{buschoff2023have,zhou2023mental}) task with only two cubes has not been thoroughly investigated. Additionally the majority of reports feature only a reduced, one dimensional decision space with respect to the stability. In the assessment of stability participants have been found to have a bias towards safety - stacks of cubes which are barely stable in a static environment are deemed unstable. The prevailing hypothesis for this bias is a security margin, separating the area deemed safe or stable from the area yielding collapsing towers (\cite{samuel2011object, zhou2023mental, liu2024object}). Since reports are based on the reduced decision space it is interesting to explore how this explanation transfers to more realistic settings.

In this paper an experiment is conducted to cover the fundamental setting. The experiment takes place in a full 3D space and features a 2D decision surface by utilizing virtual reality. 

\section{Related work}

In the field of intuitive physics there are many research directions. Most importantly in the context of this work is the study of physical support and stability in general. Most of the work in this domain focuses on the general performance of humans in detecting certain physical events. They then proceed to build models to match their performance or try to explain the biases observed in the participants. However, analyses are most often performed on aggregations of participants and not on the individual level. Hence, inter-individual differences are masked.\\
Interestingly these studies rely on cube towers, while the work which focus on (individual) behavior rather than the performance base their analyses on (one-dimensionally) tilted objects.

\paragraph{Samuel and Kerzel, 2011 \cite{samuel2011object}}
conducted two psychometric studies regarding the stability of asymmetric 2D objects. In the first experiment participants had to rotate randomly generated polygons such that \textit{it had an equal probability of falling to either side}.
Each object was composed of eight vertices which form a smaller top part and a larger bottom, resulting in a concave shape. The object was then placed on its downward tip with a randomly generated orientation. The participants were not constrained in time and no feedback has been provided regarding their accuracy.\\
In the context of our work, the second experiment is closer related. It consisted of three within-subject tasks. The first task was a forced-choice task regarding the falling direction of randomly generated objects (left, no falling, right). For the second task, the physical law of how stability can be judged in a static environment is explained to each participant. After that, participants had to judge the same set of configurations. In the final task a subset of the stimuli used in task 1 and 2 was taken, on which participants had to locate the COM.
In task 1 and 2, participants were constrained to 3 seconds. Overall accuracy was stressed and again no feedback on performance has been given.

Their analysis showed a bias towards safety, so that barely stable objects would be deemed unstable. This bias reduced after the physical law was explained with the strongest reduction in participants with greater biases. Interestingly, the entirety of the remaining bias does not seem to be sourced in inaccuracies on the perceived COM location (as investigated in task 3). Considering that the subjective location of the COM was slightly biased \textit{against} the falling direction, the bias towards safety in the judgment of stability seems even more striking.
Following the argumentation of the authors, this can be attributed to a security margin; It is safer to consider a barely stable object as unstable than the opposite. It also became apparent, that wider objects are subjected to a greater bias compared to narrower ones. This also favors the explanation, as this can be related to the effect of implied weight and consequently danger.

\paragraph{Zhou et al., 2023 \cite{zhou2023mental}} focus on the question of physical support or, more specifically, the question of how humans assess that one object supports another. For this, they propose the Counterfactual Simulation Model (CSM). It is based on the idea that humans construct a mental model of the world and simulate it forward in time, similar to a game engine. This is also called the Intuitive Physics Engine or IPE (\cite{battaglia2013ipe}). For inference it considers three independent sources of noise to account for the uncertainty humans are concerned with. To validate the CSM, three experiments are conducted in a between-subjects design. Based on this the physical judgment of humans, the CSM and a statistical model is compared. The judgment of the human participants are aggregated and the statistical model is a logistic regression model of visually computable features.\\
The experiments each feature a stack of bricks on top of a table. Given an image of the stack, the participants have to answer questions regarding the stability/support of bricks in a scene, in which a specific block has been removed from the stack.\\
In these experiments the CSM behaves similar to the aggregation of the participants while the logistic regression model does not capture participants judgments well.

\paragraph{Liu, Ayzenberg, and Lourenco, 2024 \cite{liu2024object}} aimed to find evidence, that and which visual/geometric features are sufficient to explain human judgment in intuitive physics based on the stability judgment. Thereby this paper challenges the IPE hypothesis of \cite{battaglia2013ipe}.\\
For this the authors conducted four experiments and compared the performance of adults, children and statistical models. The models either use geometric features like the centroid or the height of the object or used features extracted by a pretrained ResNet-50-network (\cite{resnet2016}). The experiments all involve a forced-choice task, asking whether a tilted object will fall on a green mat or on the red table. Note that the red table forms the support of the object. The objects in question are rendered three dimensional, placed on their lower tip and rotated around this tip by a randomly sampled angle $\theta$. Practice trials with corrective feedback preceded each experiment. Each experiment features some invariances regarding the geometric properties of the objects shown to the participants.

Their data show a distinct bias for both adults and children, such that objects which would barely stay on the table are considered falling to the mat on the ground. The bias is reported to be greater in children than in adults.
Interestingly, all statistical models the authors used have a bias towards the opposite direction and with a lower magnitude.
Their analysis suggests that humans favor the geometric center of the object as a feature to judge the falling direction of a tilted object. They found some differences between the accuracy of adults and children, for which they hypothesize a lack of precision when locating the geometric center of objects. 
Interestingly, no combination of geometric features was able to explain the totality of the bias in the responses.

\subsection{Task}

\begin{figure}
    \centering
    \includegraphics[width=0.8\linewidth]{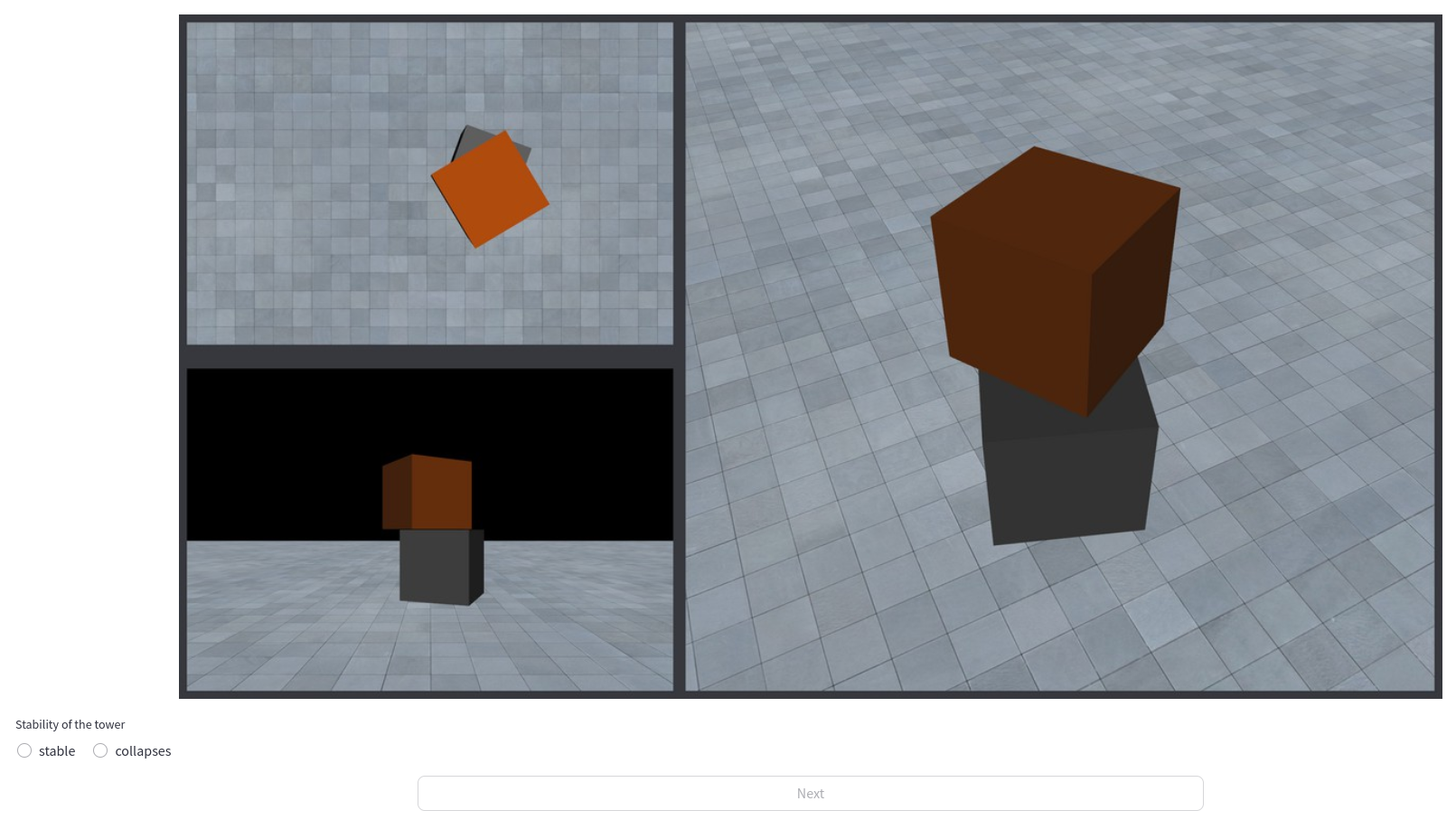}
    \caption{Example trial of the survey. You can see three perspectives: one directly from above, one lateral view and one dynamic perspective from an angle which shows the area of overlap of the cubes. It changes position and angle depending on the positions of the cubes. Below the images the stability options and the button to submit the assessment can be seen.}
    \label{fig:methods-experiment_ui}
\end{figure}

In our study the participants are tasked in a binary forced choice task to assess whether a stack of two unit-cubes collapses. The specific instruction was to judge, \textit{whether the top cube touches the floor if one were to build the tower in real life}. An example of what is presented to the participant is shown in \cref{fig:methods-experiment_ui}. It depicts three images showing the respective stack of cubes from three different perspectives. To give an assessment, the participant has to select one of the options placed directly below the images and submit via pressing the \textit{next}-button. After submitting a trial the next configuration is shown, so there is no time restriction for annotation. The stimuli shown are the same for all participants, however, in a randomized order.

We conducted an online-study using \href{clickworker.de}{clickworker.de} and obtained data from N=33 participants\footnote{we obtained data from 43 participants. However, we rejected 10 of them due to poor data quality}, with $\mathrm{age}\in [19, 66] (\mu \approx 40.24, \sigma \approx 11.37)$ and sex $\mathrm{M/W/n.s} = 25/5/1$. 

\subsection{Data}

In the stimulus set the positions and rotations of each cube is chosen randomly, such that approximately half of the stacks collapse. To achieve this, the Y coordinate (height) is fixed so the upper cube lays directly on top of the lower cube. This way there is no dynamic movement introduced through a fall from a certain height. The position in the X/Z plane of each cube is sampled from $\mathcal{N}(0, \frac{1}{2\sqrt{2}})$ such that approximately half of the towers collapse. The rotation is uniformly chosen from the interval $[0, 360)$. The whole set consists of 300 trials.

\begin{figure}
    \centering
    \includegraphics[width=\linewidth]{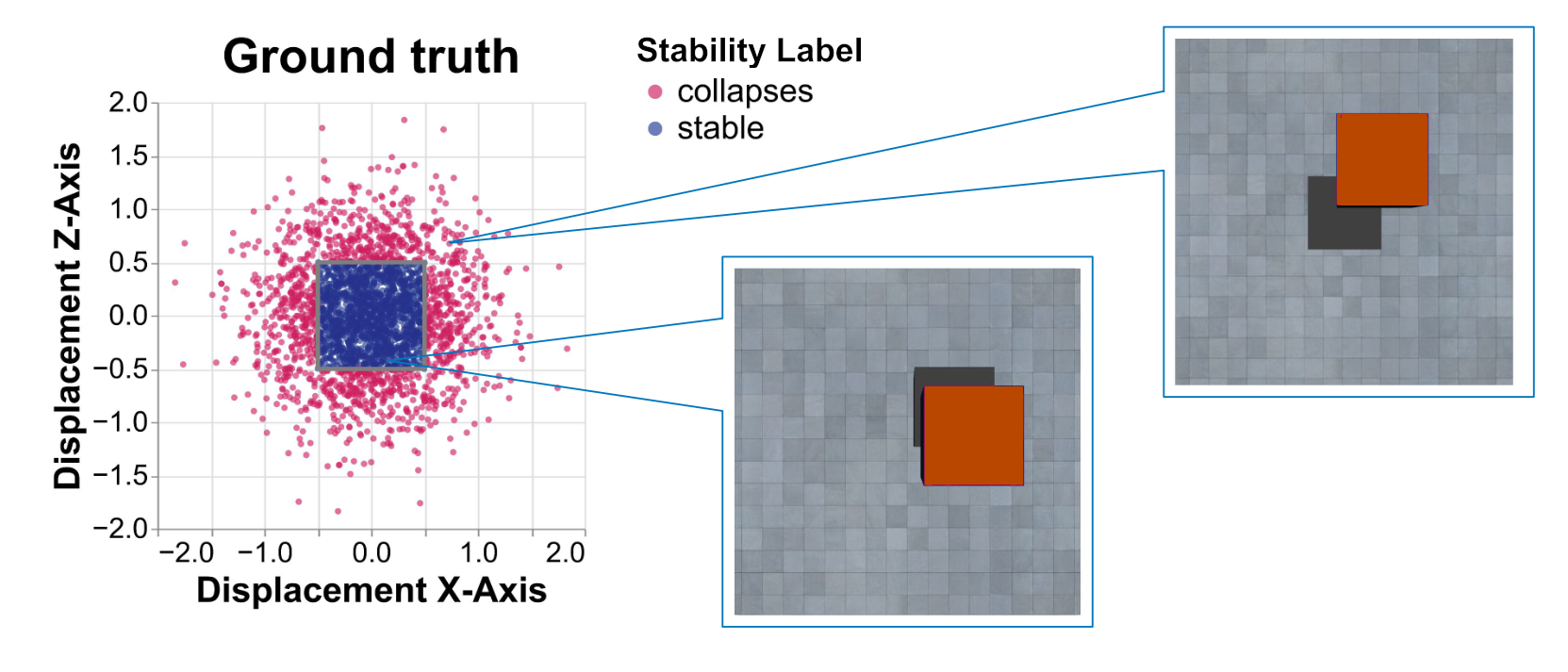}
    \caption{Position of the upper cube projected into the coordinate system of the lower cube. Each position is first projected into the lower left corner and then mirrored along the symmetry-axes of a cube in the 2D plane. The gray square depicts the outline of the lower cube.}
    \label{fig:methods-experiment_gt}
\end{figure}

In \cref{fig:methods-experiment_gt} you an see the positions of the geometrical center\footnote{Which coincides with the center of mass in this experiment} of the upper cube projected into the coordinate system of the lower cube. Note that the positions are first projected into the lower left corner of the cube and then mirrored along the symmetry-axes of a cube in the 2D plane. The gray square depicts the outlines of the lower cube. As made visible the stack remains stable, as long as the position is within bounds of the lower cube. 

For each trial we record four (multivariate) features:

\begin{itemize}
    \item \textbf{Reaction time} - duration from the onset of the stimulus to the assignment of the label
    \item \textbf{Human label} - label assigned by the human. Either of \textit{stable} and \textit{collapses}
    \item \textbf{Position and rotation of lower cube}
    \item \textbf{Position and rotation of upper cube}
\end{itemize}

\section{Results and discussion}

\begin{figure}
    \centering
    \begin{subfigure}{0.4\linewidth}
        \includegraphics[width=\linewidth]{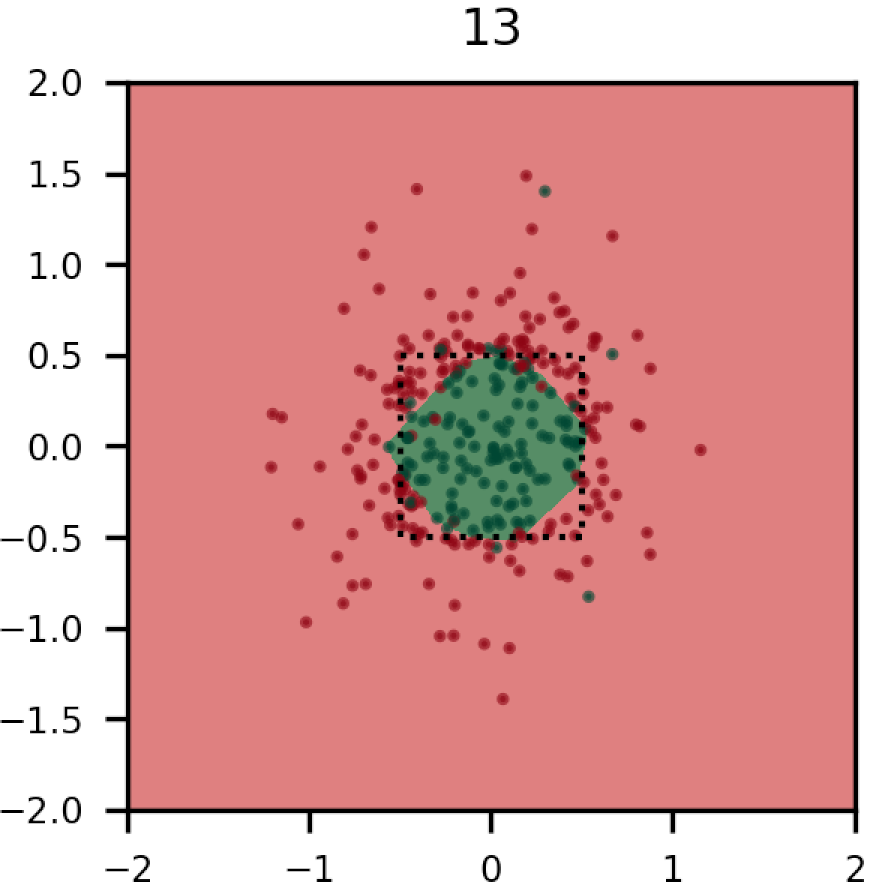}
        \caption{Rotated square 1}
        \label{fig:results-surface_approximation_samples-rot_square_1}
    \end{subfigure}
    \begin{subfigure}{0.4\linewidth}
        \includegraphics[width=\linewidth]{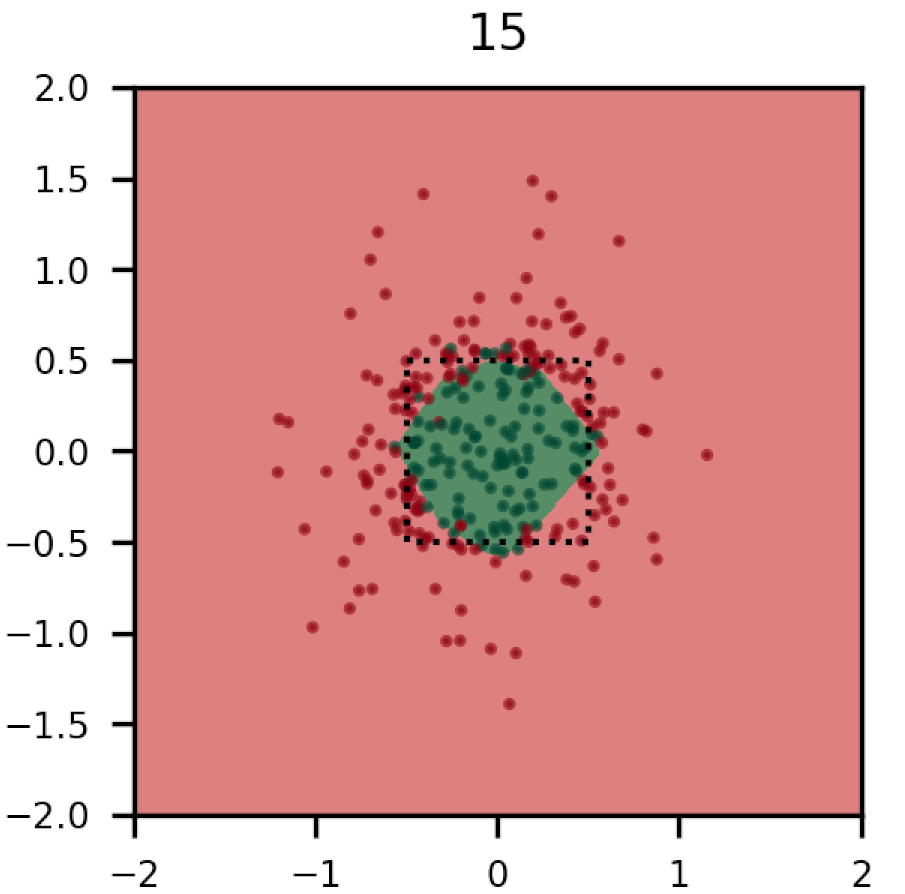}
        \caption{Rotated square 2}
        \label{fig:results-surface_approximation_samples-rot_square_2}
    \end{subfigure}

    \begin{subfigure}{0.4\linewidth}
        \includegraphics[width=\linewidth]{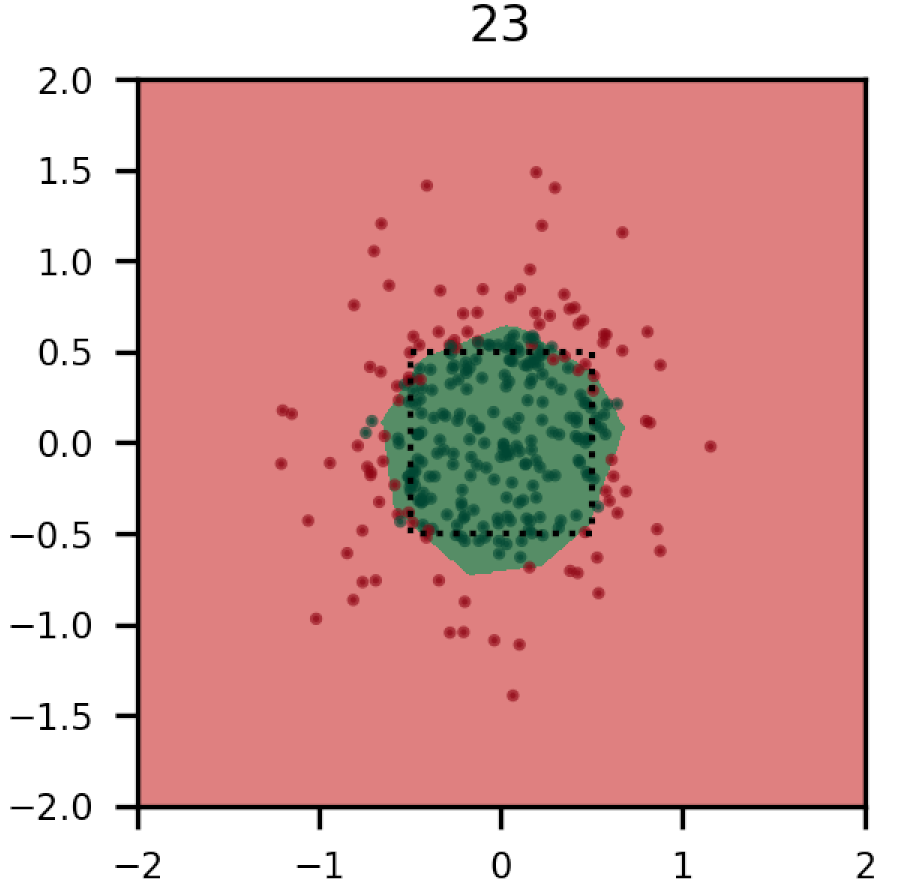}
        \caption{Wild example}
        \label{fig:results-surface_approximation_samples-wild_shape}
    \end{subfigure}
    
    \caption{Approximation of the perceived area of support}
    \label{fig:results-surface_approximation_samples}
\end{figure}

For our analysis we approximated the decision surface based on the displacement of the cubes using a multi layer perceptron with 2 hidden layers of 10 neurons each. \Cref{fig:results-surface_approximation_samples} shows examples for the approximation. The dotted, black line denotes the outlines of the lower cube. 

Our results indicate that the perceived area of support forms a rotated square. \Cref{fig:results-surface_approximation_samples-rot_square_1} and \cref{fig:results-surface_approximation_samples-rot_square_2} show examples for this. As a result, most participants classify stable configurations towards the corners of the lower cube as unstable.
Interestingly, the majority of participants who correctly assess the corner configurations as stable, make errors towards the center of the respective border, again forming a shape similar to a rotated square or circle. An example for this is shown in \cref{fig:results-surface_approximation_samples-wild_shape}.
Besides that, we observed that in almost every participant the probability of assessing a stable configuration as unstable grew anti-proportionally to the distance to the nearest corner. This also includes the bias towards safety reported by other works (e.g. \cite{samuel2011object, zhou2023mental, liu2024object}). See \cref{app:approximation-results} for visualizations of all participants. This includes the ground truth, an idealized expected participant using a safety margin as heuristic and an idealized participant with a rotated square heuristic. Naturally the participants differed in safety measures which translates into different sizes of the rectangles/shapes.

Our results show that in the realistic 3D setting participants display a more complex decision behavior which is not explainable by a simple security margin hypothesized based on 2D settings (\cite{samuel2011object, zhou2023mental, liu2024object}). One possible explanation for the observed pattern in participants' judgments is the way people naturally think about physical stability. Instead of considering a stack of cubes as a \textit{static} structure, participants may have assessed it from a \textit{dynamic} perspective. For example, when the top cube rests on the corner of the lower cube, it may be technically stable, but it is more likely to fall if disturbed, as it is vulnerable to forces from more directions. Exploring this idea could be a valuable direction for future research.

\section{Funding}
This work has been supported by the German Research Foundation DFG, as part of Collaborative Research Center (Sonderforschungsbereich) 1320 Project-ID 329551904 "EASE - Everyday Activity Science and Engineering", University of Bremen (\href{http://www.ease-crc.org/}{http://www.ease-crc.org/}). The research was conducted in subproject H01 "Sensorimotor and Causal Human Activity Models for Cognitive Architectures".

\clearpage

\bibliographystyle{unsrt}
\bibliography{references}

\clearpage

\appendix

\section{Area of support approximation} \label{app:approximation-results}

\begin{figure}[h!]
    \centering
    \includegraphics[width=\linewidth]{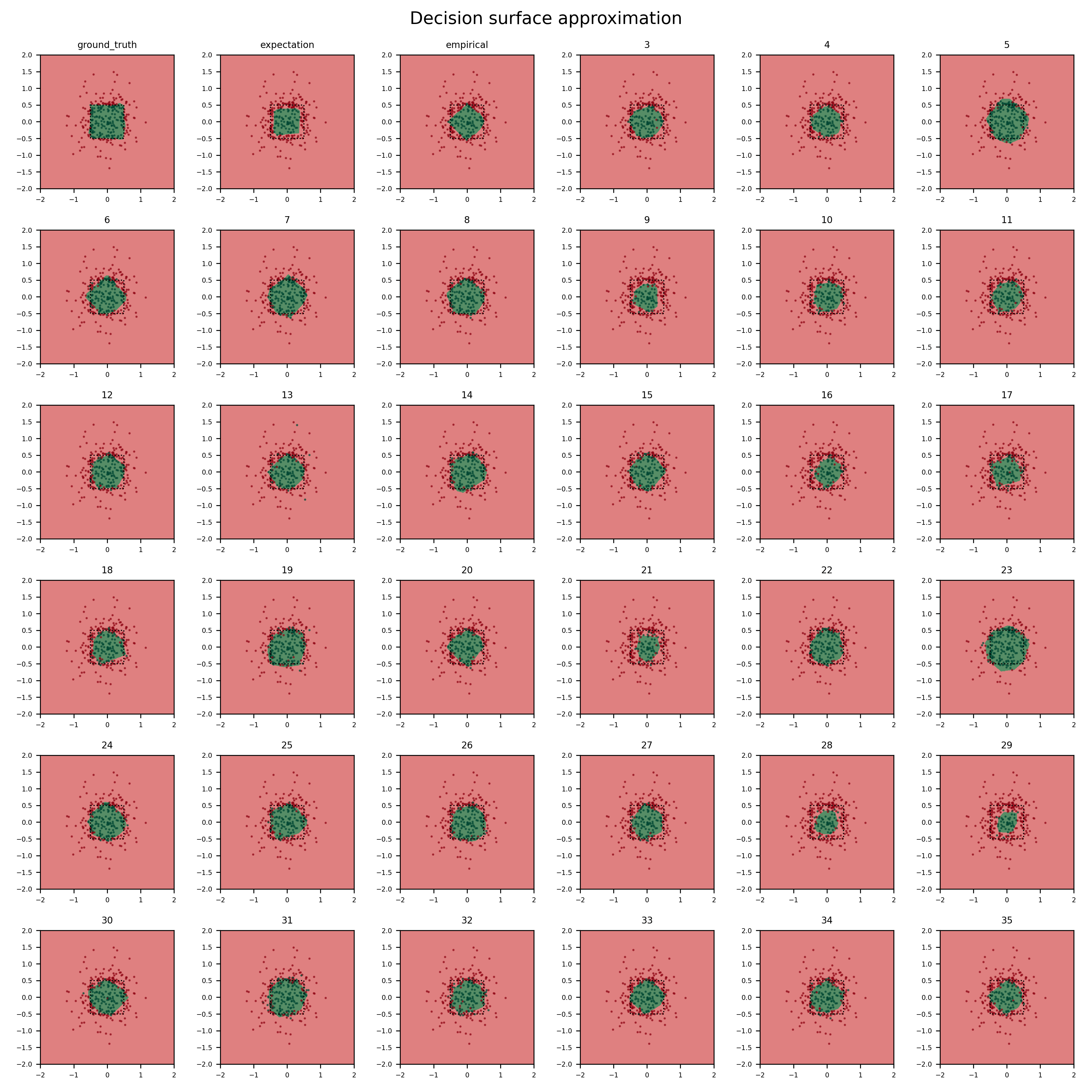}
    \label{fig:app-approximation-results}
\end{figure}

\end{document}